\documentclass[twocolumn,amsmath,amssymb,longbibliography,noeprint]{revtex4-1}

\usepackage{bbm}
\usepackage[caption=false]{subfig}
\usepackage{overpic}
\usepackage{amsmath}
\usepackage{epsfig,psfrag}
\usepackage{pstricks}
\usepackage{dcolumn}
\usepackage{bm}
\usepackage{graphicx}
\usepackage{enumitem}   
\usepackage{color}
\usepackage{version}
\usepackage{hyperref}
\definecolor{darkblue}{RGB}{46,48,146}
\definecolor{codered}{RGB}{214,99,56}
\definecolor{codegreen}{RGB}{86,160,24}
\definecolor{codegray}{RGB}{150,150,150}
\definecolor{codepurple}{RGB}{197,126,126}
\hypersetup{
   colorlinks=true,
   citecolor={darkblue},
   linkcolor={darkblue},
   urlcolor={darkblue},
   pdfhighlight=/I
}

\usepackage{times}
\usepackage{xcolor}
\usepackage{braket}

\begin{document}

\title{
Braiding by Majorana Tracking and Long-Range CNOT Gates with Color Codes}
\author{Daniel Litinski and Felix von Oppen}
\affiliation{Dahlem Center for Complex Quantum Systems and Fachbereich Physik, Freie Universit\"at Berlin, Arnimallee 14, 14195 Berlin, Germany}

\begin{abstract}
Color-code quantum computation seamlessly combines Majorana-based hardware with topological error correction. Specifically, as Clifford gates are transversal in two-dimensional color codes, they enable the use of the Majoranas' nonabelian statistics for gate operations at the code level. Here, we discuss the implementation of color codes in arrays of Majorana nanowires that avoid branched networks such as T-junctions, thereby simplifying their realization. We show that, in such implementations, nonabelian statistics can be exploited without ever performing physical braiding operations. Physical braiding operations are replaced by Majorana tracking, an entirely software-based protocol which appropriately updates the Majoranas involved in the color-code stabilizer measurements. This approach minimizes the required hardware operations for single-qubit Clifford gates. For Clifford completeness, we combine color codes with surface codes, and use color-to-surface-code lattice surgery for long-range multi-target CNOT gates which have a time overhead that grows only logarithmically with the physical distance separating control and target qubits. With the addition of magic state distillation, our architecture describes a fault-tolerant universal quantum computer in systems such as networks of tetrons, hexons, or Majorana box qubits, but can also be applied to non-topological qubit platforms.

\end{abstract}

\maketitle

\section{Introduction}\label{sec:intro}

The scalable fabrication of high-fidelity qubit platforms is necessary for large-scale quantum computation. In topological quantum computing~\cite{Kitaev2003,Nayak2008,TerhalRMP}, Majorana-based architectures~\cite{Fu2008,Oreg2010,Lutchyn2010,Alicea2012,Beenakker2013} have been proposed as candidates for such high-fidelity qubits. Among the advantages that Majorana-based qubits may offer in comparison to conventional qubits are long coherence times, high-fidelity single-qubit Clifford gates by braiding, and ancilla-free stabilizer measurements for quantum error correction. Recent experiments have demonstrated considerable progress towards realizing Majorana zero modes (Majoranas) in topological superconductors~\cite{Mourik2012,Das2012,Albrecht2016a,Deng2016,Suominen2017}, but topological qubits are yet to be implemented and their advantages remain to be confirmed experimentally.

As Majorana-based qubits are still expected to have a finite lifetime~\cite{VanWoerkom2015,Higginbotham2015,Albrecht2016,Rainis2012}, quantum error correction is necessary for fault-tolerant quantum computation~\cite{Preskill1998,Bravyi2010,Vijay2015,Vijay2016,Landau2016,Plugge2016,Vijay2017,Hastings2017}. In a recent work~\cite{Litinski2017}, we have argued that Majorana-based qubits are particularly well-suited for quantum error correction with topological color codes~\cite{Bombin2006,Landahl2011}. Unlike in surface codes~\cite{Fowler2012,TerhalRMP,Campbell2016}, the Clifford gates are transversal in two-dimensional color codes. Thus, logical Clifford gates are implemented on the code level by performing independent Clifford gates on all (pairs of) physical qubits that make up the logical qubit(s). Importantly, this transversal gate set enables the use of braiding for logical gates, thereby fully exploiting the topological protection provided by Majorana-based hardware. Moreover, the existence of transversal gates has additional advantages. Independent operations on the physical qubit do not spread errors during gate operations and minimize the time overhead by allowing parallel implementation. 

We discussed a physical implementation of a Majorana color code which relies on topological superconductor networks, where Majoranas are braided by moving them through branched geometries. Moving Majoranas was also necessary for lattice surgery~\cite{Horsman2012,Landahl2014} and magic state distillation~\cite{Bravyi2005}, which are required to complete the universal gate set. However, current experiments where proximity-coupled quantum wires are driven into the topological phase require an external magnetic field in the direction of the wire. This might constitute a significant obstacle for all braiding protocols that rely on the movement of Majoranas in branched geometries~\cite{Alicea2011,VanHeck2012,Hyart2013,Aasen2016}. Moreover, movement of Majoranas has also been shown to be susceptible to thermal noise~\cite{Pedrocchi2015}. To overcome these problems, recent works have proposed architectures that avoid T-junctions, and are instead based on arrays of parallel nanowires~\cite{Landau2016,Plugge2016,Plugge2016a,Vijay2016a,Karzig2016}. In this work, we show that Majorana color codes can be naturally implemented in such setups, thereby entirely avoiding T-junctions and explicit movement of Majoranas. 

An important aspect of the recent works on implementing topological qubits in arrays of parallel wires is that braiding is no longer performed by moving Majoranas or coupling Majoranas in judicious ways, but instead by measuring a sequence of two-Majorana parity operators~\cite{Karzig2016,Bonderson2008}. Thus, these architectures shift the experimental challenge away from the fabrication of branched geometries, and towards the access to measurements of various local Majorana parity operators. It was pointed out that in this setting, single-qubit Clifford gates can be implemented by an entirely classical software-based procedure~\cite{Hastings2015,Clarke2016,Karzig2016}, which we refer to as \emph{Majorana tracking}. This procedure obviates explicit hardware operations and, similar to Pauli tracking, only requires appropriate updates of the qubit's reference frame. We show that the transversal gates of color codes take Majorana tracking to the level of logical qubits, and thereby reduce the overhead of logical single-qubit Clifford gates to a (classical) minimum. The only required hardware operation is the measurement of certain local Majorana parity operators corresponding to the stabilizers of the quantum error-correcting code.

Universal fault-tolerant quantum computation can be achieved by implementing two more logical gates: the controlled-NOT (CNOT) gate and the $T$ gate. While logical CNOTs can in principle be implemented transversally using physical CNOTs, this requires nonlocal physical gates. Instead, it is more convenient to implement logical CNOTs via lattice surgery~\cite{Horsman2012}, which only requires local operations. Here, we present a scheme for logical CNOTs which combines  color codes with surface code ancillas and employs color-to-surface-code lattice surgery~\cite{Nautrup2016}. This protocol implements the CNOT gate without the need for any movement of Majoranas while retaining the long-range communication between color code qubits of our earlier implementation. This also  provides us with a long-range multi-target CNOT, which is an essential part of magic state distillation protocols, thereby completing the universal gate set.

Here, we are mainly interested in providing a proof-of-principle implementation of Majorana-based color-code quantum computation in a network of tetrons~\cite{Karzig2016}. However, it should be emphasized that the combination of Majorana-based hardware with color code error correction transcends our specific implementations. It seems likely that this combination can be used to one's advantage in many, if not all future implementations of fault-tolerant Majorana-based quantum computation. In fact, our lattice-surgery-based scheme can in principle be applied even to non-topological qubit architectures. Nevertheless, the robustness of physical single-qubit Clifford gates and the ease of stabilizer measurements are key advantages of Majorana-based qubit platforms.

\section{Majorana Tracking and Color Codes}
\label{sec:nanowirearrays}

A Majorana-based qubit can be defined using three Majorana fermions $\gamma_1$, $\gamma_2$, and $\gamma_3$ with $\{\gamma_i,\gamma_j\} = 2\delta_{i,j}$ and $\gamma_i = \gamma_i^\dagger$. Since Majorana fermions in physical systems always come in pairs, it is convenient to define the qubit using four Majoranas with fixed total parity $-\gamma_1\gamma_2\gamma_3\gamma_4 = 1$, such that all two-Majorana parity operators $i\gamma_m\gamma_n$ of a qubit can be expressed in terms of the first three Majoranas. In the Schr\"odinger picture, a qubit is defined using two computational states $\ket{0}$ and $\ket{1}$ in the $\sigma_z$-basis . For our purposes, it will be instead more useful to express the qubit in the Heisenberg picture, where we define the qubit by its $\sigma_x$- and $\sigma_z$-Pauli operators, which in their default state are
\begin{equation}
	\sigma_z = i\gamma_1\gamma_2, \, \, \sigma_x = i \gamma_2 \gamma_3 \, .
\label{eqn:tracking1}
\end{equation}
Consequently, the remaining Pauli operator is $\sigma_y = i\gamma_1\gamma_3$. One can check that these operators square to unity and fulfill the commutation relations of Pauli operators $[\sigma_i,\sigma_j]=2i\epsilon_{ijk}\sigma_k$.
Two Pauli operators are sufficient to define a qubit, as any single-qubit unitary operator can be expressed in terms of $\sigma_x$ and $\sigma_z$ via the Euler decomposition. 

The basic framework of our architecture are nanowire arrays, which are two-dimensional networks of Majorana-based physical qubits. Several proposals for implementations of such nanowire arrays can be found in Refs.~\cite{Landau2016,Plugge2016,Plugge2016a,Vijay2016a,Karzig2016}. Based on these proposals, we assume that the following basic operations can be implemented in the nanowire array:
\begin{enumerate}[label=(\roman*)]
	\item Measurements of local $2n$-Majorana fermion parity \linebreak operators $i^n \prod_{i=1}^{2n}\gamma_i$
	\item Some non-robust implementation of a possibly faulty $T$ gate ($\pi$/8 gate) on physical qubits
\end{enumerate} 
As already emphasized above, we do \textit{not} require that the Majoranas can be moved through the network.

\begin{figure}
\centering
\def\svgwidth{\linewidth}
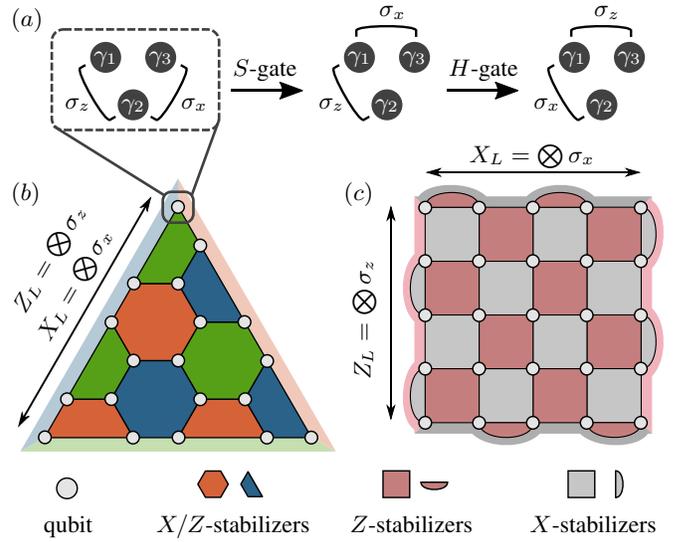
\caption{$(a)$ Majorana-based qubit consisting of three Majoranas and an example for Majorana tracking. Starting from the default encoding $\sigma_z=i\gamma_1\gamma_2$ and $\sigma_x=i\gamma_2\gamma_3$, an $S$ gate changes the encoding to $\sigma_z=i\gamma_1\gamma_2$ and $\sigma_x = i\gamma_1\gamma_3$. A subsequent $H$ gate changes it to $\sigma_z=i\gamma_1\gamma_3$ and $\sigma_x=i\gamma_1\gamma_2$. Keeping track of the current encoding for each physical qubit via a classical computer is referred to as Majorana tracking. $(b)$ Triangular color code qubits can be defined on a hexagonal lattice, where each vertex is a Majorana-based qubit comprised of three Majoranas (or four Majoranas with fixed total parity). Each face corresponds to two stabilizers $\sigma_z^{\otimes m}$ and $\sigma_x^{\otimes m}$. Products of $\sigma_z$ and $\sigma_x$ operators along any one of the three boundaries correspond to logical $Z_L$ and $X_L$ operators. $(c)$ In surface code qubits, on the other hand, the support of $X$- and $Z$-stabilizers does not coincide, and the two different edges correspond to the $Z_L$ and $X_L$ operators, respectively.}
\label{fig:qubits}
\end{figure}

\subsection{Physical single-qubit Clifford gates: Majorana Tracking}

The first operation includes the measurement of all two-Majorana parity operators~--~and therefore all Pauli operators~--~of a physical qubit. This enables the use of Majorana tracking for a particularly simple implementation of the single-qubit Clifford gates as pioneered in Refs.~\cite{Hastings2015,Karzig2016}. These gates map Pauli operators onto other Pauli operators and are products of Hadamard ($H$) and phase ($S$) gates. Specifically, the action of these two gates on the Pauli operators is
\begin{equation}\begin{split}
H:\qquad \sigma_z \rightarrow \sigma_x&\, , \quad \sigma_x \rightarrow \sigma_z \, ,\\ 
S:\qquad \sigma_z \rightarrow \sigma_z&\, , \quad \sigma_x \rightarrow i\sigma_x\sigma_z \, .\end{split}\end{equation}
Since the $H$ and $S$ gates can be implemented by braiding, their application simply redefines the Majoranas involved in the corresponding two-Majorana parity operator. Thus, instead of physically braiding Majoranas, one can alternatively keep track of the Majorana operators that define the $\sigma_z$ and $\sigma_x$ of each physical qubit using a classical computer.
In analogy to Pauli tracking~\cite{PauliTracking}, we refer to this procedure as Majorana tracking.

As a concrete example, consider the sequence of operations shown in Fig.~\ref{fig:qubits}a. Starting from the default encoding in Eq.~\eqref{eqn:tracking1}, an $S$ gate takes the encoding to $\sigma_z=i\gamma_1\gamma_2$ and $\sigma_x=i\gamma_1\gamma_3$. A subsequent $H$ gate will exchange these two operators to  $\sigma_z=i\gamma_1 \gamma_3$ and $\sigma_x=i\gamma_1\gamma_2$. So instead of initializing the qubit in a $\sigma_z$-eigenstate, physically performing the two gates and then reading out the qubit in the $\sigma_z$-basis, one can simply initialize the qubit in a $\sigma_z$-eigenstate and then measure the $\sigma_y=i\gamma_1\gamma_3$ operator.

It should not be surprising that Clifford gates can be treated entirely classically, as these gates can be efficiently simulated on a classical computer by virtue of the Gottesman-Knill theorem~\cite{Gottesman1999}. As this classical tracking of Pauli operators can also be done with non-topological, e.g. superconducting, qubits, it would appear that Majorana tracking does not utilize braiding. However, conventional qubits still require a hardware operation for the rotation of the Pauli basis during readout. While for conventional qubits the angle of rotation is susceptible to errors, with Majorana tracking the angle is robust. Even though Majorana tracking eliminates any hardware operation for single-qubit Clifford gates, it leads to the same robust gates as braiding. In this sense, Majorana tracking \textit{is} braiding.

Therefore, Majorana tracking can also be used to probe the nonabelian statistics of Majorana zero modes. With Majorana tracking, a fusion-rule detection experiment in the spirit of Ref.~\citep{Aasen2016} would correspond to alternating measurements of $\sigma_z$ and $\sigma_x$. If Majorana zero modes are present, the measurement results will be entirely uncorrelated, whereas repeated measurements of $\sigma_z$ will always yield the same result. In this way, the fusion-rule detection experiment probes the robustness of the single-qubit Clifford gates.

\subsection{Logical single-qubit Clifford gates: color codes}
\label{sec:logcliff}

The gates that can be implemented by Majorana tracking are physical gates on physical qubits. However, these Majorana-based qubits only have a finite lifetime which is set by processes that introduce errors, such as quasiparticle poisoning. In order to quantum compute beyond the coherence time of physical qubits, a quantum error-correcting code needs to be used. This allows for \textit{fault-tolerant} quantum computing, which relies on combining many physical qubits into one logical qubit. This not only replaces the physical error rate by an (in principle) arbitrarily low logical error rate, but also substitutes physical gates with logical gates.

It is desirable to use Majorana tracking for logical gates in order to minimize the overhead of single-qubit Clifford gates. But this can only be done if these gates are transversal gates of the error-correcting code, i.e., if the logical $H$ and $S$ gates are $H_L = H^{\otimes n}$ and $S_L = S^{\otimes n}$, where $n$ is the number of physical qubits in a logical qubit. This is precisely the reason why color codes are a natural choice for Majorana-based qubits~\cite{Litinski2017}, as their set of transversal gates are the Clifford gates.

\begin{figure*}
\centering
\def\svgwidth{\linewidth}
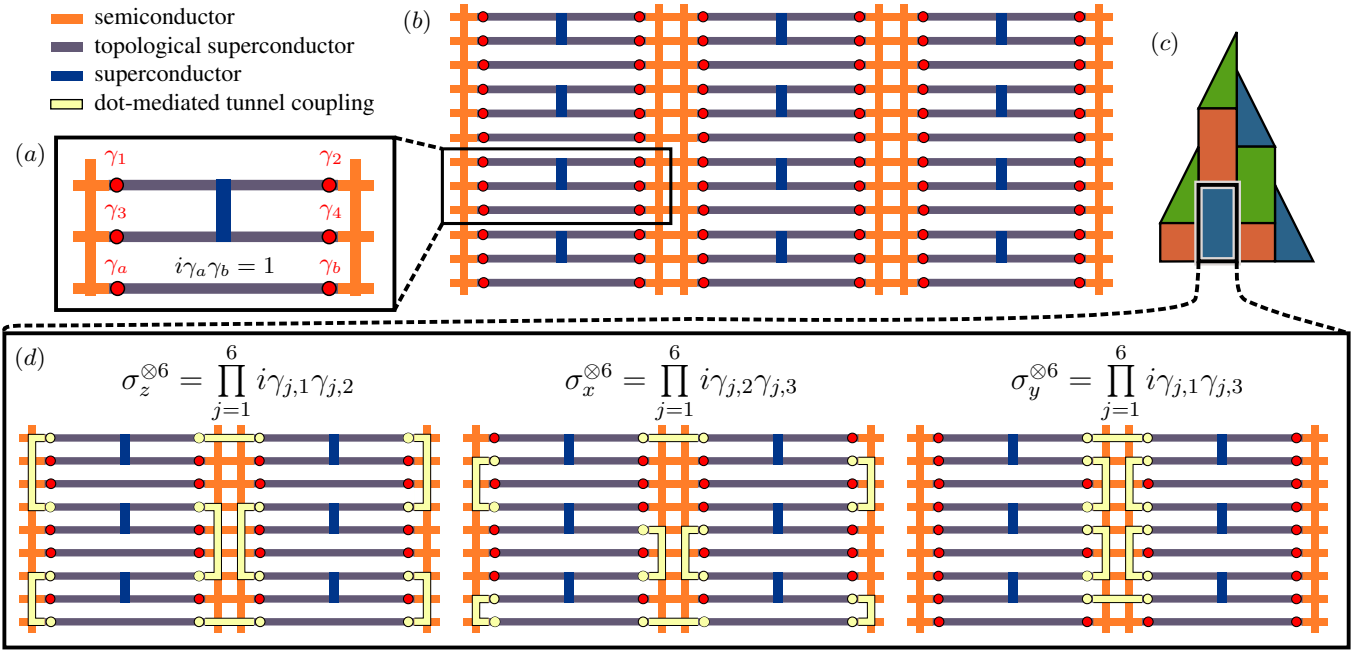
\caption{$(a)$ A single tetron~\cite{Karzig2016} consists of two topological superconducting nanowires hosting four Majoranas $\gamma_1\dots\gamma_4$. The two wires are bridged by an ordinary superconductor which fixes the total parity sector $-\gamma_1\gamma_2\gamma_3\gamma_4$. In addition, a coherent link formed by a topological superconducting nanowire hosting Majoranas $\gamma_a$ and $\gamma_b$ with a fixed parity is part of the basic building block. The three Majorana nanowires are connected to a semiconducting nanowire network via gate-tunable tunnel couplings. $(b)$ A network of tetrons forms a square lattice of physical qubits. $(c)$ In such a square lattice, a triangular color code qubit can be defined in a brick wall geometry. $(d)$ Configurations of the tunnel couplings used to measure three different stabilizers, which are either products of $\gamma_1\gamma_2$ of each tetron, or $\gamma_2\gamma_3$ , or $\gamma_1\gamma_3$. One can verify that in all three cases, the circular paths only contain the corresponding Majoranas of each tetron, and Majoranas that belong to coherent links.}
\label{fig:tetrons}
\end{figure*}

Using triangular color codes~\cite{Bombin2006,Landahl2011}, a logical qubit is encoded by $n$ physical qubits located at the vertices of the triangle with hexagonal tiling shown in Fig.~\ref{fig:qubits}b. The figure shows a specific qubit with code distance $d=5$, but this construction can be generalized to arbitrary odd code distances. As described above, each physical qubit effectively corresponds to three Majorana fermions. The logical qubit is initialized in the logical $\ket{0_L}$-state by initializing the $n$ physical qubits in the $\ket{0}$-state by measuring $i\gamma_1\gamma_2$ of all physical qubits, and then measuring the stabilizers of the code. These $n-1$ stabilizers are defined by the faces of the hexagonally tiled triangle, with each face defining an $X$-type stabilizer $\mathcal{O}_X = \sigma_x^{\otimes m}$ and a $Z$-type stabilizer $\mathcal{O}_Z = \sigma_z^{\otimes m}$, where $m$ is the number of qubits that are part of a face. In analogy to Majorana surface codes~\cite{Landau2016,Plugge2016}, one can represent color codes as Majorana fermion codes by identifying $\sigma_z = i\gamma_1\gamma_2$ and $\sigma_x=i\gamma_2\gamma_3$. Thus, the stabilizers in Fig.~\ref{fig:qubits}b are products of 8 or 12 Majorana fermions. A color code qubit can be read out in any Pauli basis by measuring all physical qubits in the corresponding basis.

Quantum error-correcting codes typically operate in cycles. In each code cycle, the stabilizers are measured to determine the error syndrome, errors are corrected, and logical gate operations are performed. The single-qubit logical Clifford gates are transversal in color codes, i.e., a logical $H$ gate corresponds to physical $H$ gates on all qubits, whereas a logical $S$ gate is a combination of physical $S$  and $S^\dagger$ gates. For instance, a conventional procedure for a logical $S$ gate would be to measure and correct the error syndrome, transversally perform physical $S$ and $S^\dagger$ gates (e.g., by braiding), and again measure the error syndrome and correct errors. With Majorana tracking, the physical gate operations are replaced by an update of the $\sigma_z$ and $\sigma_x$ operators of all physical qubits. While the $\sigma_z$ operators are unaffected by the $S$ and $S^\dagger$ gates, the $\sigma_x$ operators are changed from $i\gamma_2\gamma_3$ in the default encoding to $\pm i\gamma_1\gamma_3$. In other words, Majorana tracking modifies which Majorana fermions are part of the stabilizer measurements. In the case of an $S$ gate, the $X$-type stabilizers $\sigma_x^{\otimes 6}$ are replaced by $Y$-type stabilizers $\sigma_y^{\otimes 6}$ in the following rounds of syndrome measurement, i.e., the $X$-type stabilizers are changed from products of $i\gamma_2\gamma_3$ to products of $i\gamma_1\gamma_3$. 

In this way, keeping track of the current Majorana composition of $\sigma_z$ and $\sigma_x$ for each physical qubit implements logical single-qubit Clifford gates with Majorana color codes. However, considering that in the default encoding, the measurement of $X$- and $Z$-type stabilizers automatically measures the $Y$-type stabilizers as their product, it is not necessary to actually change the measured stabilizers after the application of single-qubit Clifford gates. This is due to the fact that the support of $X$- and $Z$-stabilizers of color code qubits coincides. Instead, Majorana tracking on the level of logical qubits, similar to the tracking procedure on physical qubits, merely updates the Majoranas measured during qubit readout. In the previous example of a logical $S$ gate, tracking changes the $X_L$-basis readout from a measurement of $i\gamma_2\gamma_3$ to the measurement of $i\gamma_1\gamma_3$ of \textit{all} physical qubits. An appropriate update of certain stabilizer operators will be necessary as soon as CNOT gates are involved, because this introduces $X$- and $Z$-stabilizers whose support does \textit{not} coincide, as we discuss in Sec.~\ref{sec:lrcnot}. 

Previously~\cite{Litinski2017}, we argued that Majorana-based qubits and color codes are a natural fit, as transversal Clifford gates allow for the use of braiding for logical gates. In the context of the present work, this statement takes the equivalent form: Owing to braiding by Majorana tracking, Majorana-based qubits can be read out in every Pauli basis without the need for intermediate hardware operations. Similarly, due to transversal Clifford gates, color code qubits can be measured in every logical Pauli basis without requiring intermediate logical gates. Thus, color codes in combination with Majorana-based qubits reduce the overhead of \textit{logical} single-qubit Clifford gates to a minimum.


\section{Implementation with Tetrons}
\label{sec:colorcodes}

In this section, we present a proof-of-principle implementation of a Majorana color code in a nanowire array, which differs from our earlier setup~\cite{Litinski2017} in two essential ways. First, the present implementation relies on recent suggestions to realize Majorana-based topological qubits using only parallel topological superconducting nanowires. Second, as a consequence of implementing braiding at the code level by Majorana tracking, Majoranas no longer need to be moved within the network.

\begin{figure}[b]
\centering
\def\svgwidth{0.89\linewidth}
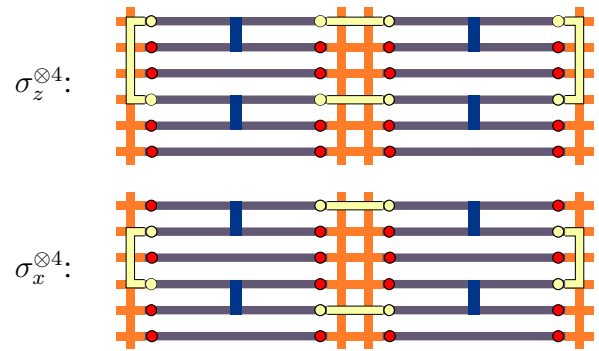
\caption{Tetron tunnel coupling configurations for the measurement of the four-qubit stabilizers of the surface code.}
\label{fig:surfacecodetetron}
\end{figure}

\begin{figure*}
\centering
\def\svgwidth{\linewidth}
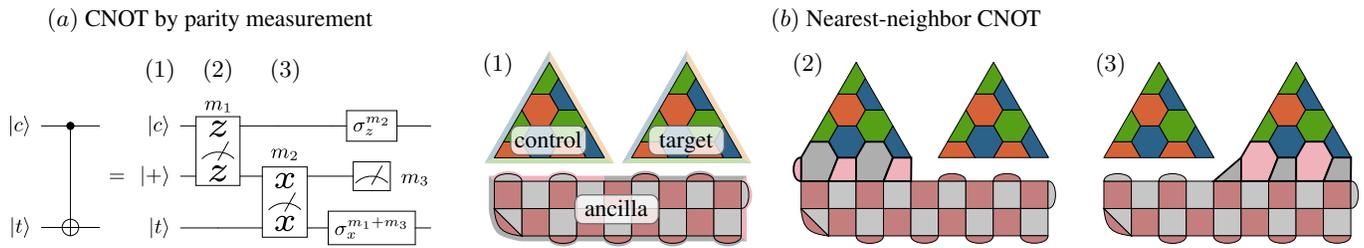
\caption{$(a)$ Quantum circuit corresponding to the logical CNOT gate between a control $\ket{c}$ and target $\ket{t}$ by lattice surgery, using an ancilla qubit initialized in the $\ket{+}$-state. First, the $ZZ$-parity $\sigma_z\otimes\sigma_z$ between control and ancilla is measured. Next, the $XX$-parity $\sigma_x\otimes\sigma_x$ between ancilla and target is measured, and the ancilla is read out in the $\sigma_z$-basis. The three measurement outcomes are used to determine a final Pauli correction. $(b)$ Nearest-neighbor CNOT between color code qubits using a surface code ancilla. Lattice surgery between the green color code boundary and the purple surface code boundary $(b2)$ measures the $ZZ$-parity, whereas surgery with the gray surface code boundary $(b3)$ constitutes an $XX$-parity measurement. 
}
\label{fig:cnot1}
\end{figure*}

Specifically, we present an implementation in a network of tetrons. A tetron~\cite{Plugge2016a,Karzig2016} is a qubit (Fig.~\ref{fig:tetrons}a) that consists of two topological superconducting nanowires with four Majoranas $\gamma_1\dots\gamma_4$ with fixed total parity $-\gamma_1\gamma_2\gamma_3\gamma_4$. The fixed parity sector not only protects the qubit from quasiparticle poisoning, but also enables the use of the fourth Majorana $\gamma_4$ for quantum computation. In the even parity sector, we can identify $\sigma_z =  i\gamma_1\gamma_2 = i\gamma_3\gamma_4$ and $\sigma_x =  i\gamma_2\gamma_3 = i\gamma_1\gamma_4$.
Furthermore, each tetron contains a third floating Majorana nanowire with Majoranas $\gamma_a$ and $\gamma_b$ and fixed parity $i\gamma_a\gamma_b$ acting as a coherent link. Gate-tunable tunnel couplings connect the three topological superconducting nanowires to a semiconductor network. The network of tetrons shown in Fig.~\ref{fig:tetrons}b corresponds to the architecture described in Ref.~\cite{Karzig2016}, but with two vertical semiconductor wires between adjacent tetrons, instead of just one. (However, this is not a requirement, as the implementation of a color code is also possible in the setup described in Ref.~\cite{Karzig2016}, see Appendix~\ref{app:singlewire}.) The tetron qubits form a square lattice which can be used to encode color code qubits in a brick wall geometry, see Fig.~\ref{fig:tetrons}c.

With tetrons, $2n$-Majorana parity operators are measured by opening tunnel couplings between tetrons such that they form a closed path, as discussed in Ref.~\cite{Karzig2016}. The semiconducting segments that couple neighboring tetrons form quantum dots. Their energy levels are shifted by virtual processes that tunnel electrons around this closed path. As these processes involve each Majorana operator along this path exactly once, the energy shift depends on the product of the Majoranas, i.e., on the $2n$-Majorana parity. Suitable spectroscopy on the dots can thus be used to measure this parity. Essentially, this measures the product of all Majoranas along a closed loop formed by the gate-tunable tunnel couplings, thereby implementing local $2n$-Majorana parity measurements. 

Consider the configuration of the tunnel couplings in the left panel of Fig.~\ref{fig:tetrons}d. The circular path formed by the coupled tetrons involves the $\gamma_1$ and $\gamma_2$-Majoranas of each tetron, and four Majoranas that belong to coherent links. Since the parity of the coherent links is known, this  configuration can be used to measure the 12-Majorana operator that corresponds to $\sigma_z^{\otimes 6}$ in the default encoding. The center panel shows a configuration that measures the product of $\gamma_2$ and $\gamma_3$-Majoranas of each tetron, corresponding to a $\sigma_x^{\otimes 6}$ operator in the default encoding. Note that since the total parity sector of each tetron is fixed, $i\gamma_2\gamma_3 = i\gamma_1\gamma_4$. It is also possible to measure $\sigma_y^{\otimes 6}$-stabilizers, whose configuration is shown in the right panel of Fig.~\ref{fig:tetrons}d and does not require the use of coherent links.

\section{Universal Quantum Computation}
\label{sec:quantumcomputation}

Having discussed logical single-qubit Clifford gates, two more gates are required for universal quantum computation: a logical controlled-not (CNOT) gate and a logical $T$ gate, where $T = \exp(i\sigma_z \pi/8)$. As discussed in Ref.~\cite{Litinski2017} for topological superconductor networks with branched geometries, these operations can be implemented by lattice surgery and magic state distillation, respectively. Here, we adapt this scheme to architectures where Majoranas cannot be moved, such that only the aforementioned stabilizer measurements and physical $T$ gates are required.

\subsection{Long-range CNOT gates}
\label{sec:lrcnot}

A logical CNOT gate between two color code qubits can be implemented using lattice surgery with the help of an ancilla qubit~\cite{Landahl2014}. This scheme effectively realizes the circuit identity shown in Fig.~\ref{fig:cnot1}a using an ancilla qubit initialized in the $\sigma_x$-eigenstate $\ket{+}$. A CNOT gate corresponds to a $ZZ$-parity ($\sigma_z \otimes \sigma_z$) measurement between the control qubit and the ancilla, a subsequent $XX$-parity measurement between ancilla and target, and a final $\sigma_z$-measurement of the ancilla qubit. Note that the protocol requires parity measurements between \emph{logical} qubits. Lattice surgery is a fault-tolerant protocol for such parity measurements requiring only the measurement of additional stabilizer operators straddling the adjacent boundaries of two logical qubits. Essentially, lattice surgery measures the product of the logical operators defined on the two boundaries.

These boundary operators depend on the kind of logical qubit that is used.
Triangular color code qubits have three boundaries: a red, a green, and a blue edge. Strings of $\sigma_z$ and $\sigma_x$ operators along any of these edges are logical $Z_L$ and $X_L$ operators, respectively, as illustrated in Fig.~\ref{fig:qubits}b. Surface code qubits, on the other hand, have pairs of opposing $X$- and $Z$-edges, also referred to as rough and smooth edges, and are drawn as gray and purple edges in Fig.~\ref{fig:qubits}c. The logical $Z_L$ operator is a product of $\sigma_z$ operators along any of the two purple boundaries, whereas $X_L$ is a string of $\sigma_x$ operators along a gray boundary. With tetrons, the measurement of the four-qubit surface code stabilizer operators $\sigma_z^{\otimes 4}$ and $\sigma_x^{\otimes 4}$ is similar to the color code stabilizer measurements, as shown in Fig.~\ref{fig:surfacecodetetron}. In the following protocols, we use surface codes instead of color codes to encode ancilla qubits, as the CNOT protocol does not require the use of any transversal Clifford gates on the ancillas.  Apart from lattice surgery, the only required surface code operations are the initialization in a $\sigma_x$-basis eigenstate, and a $\sigma_z$-basis measurement, both of which amount to $\sigma_x$- and $\sigma_z$-measurements of all physical qubits, and to stabilizer measurements. The main advantage of using surface code ancillas for CNOTs between color code qubits is that, compared to color code ancillas, they require fewer qubits and feature lower-weight stabilizers, as we discuss in Appendix \ref{app:surgery}.

\begin{figure}[b]
\centering
\def\svgwidth{\linewidth}
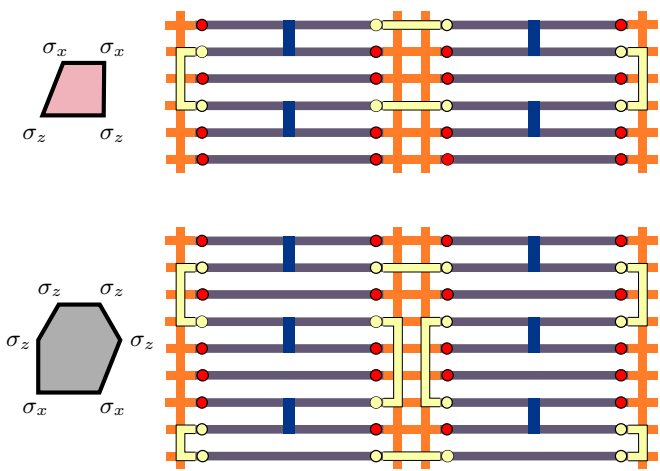
\caption{Tetron tunnel coupling configurations to measure the lattice surgery boundary stabilizers between control and ancilla qubit for the case of a preceding $H$ gate on the control qubit. The measured operators are $\sigma_x^{\otimes 2} \otimes \sigma_z^{\otimes 2}$ (light purple) and $\sigma_z^{\otimes 4} \otimes \sigma_x^{\otimes 2}$ (dark gray). This effectively describes a logical $\sigma_x \otimes \sigma_z$ measurement between control and ancilla. This protocol can be straightforwardly adapted to measure any other product of two logical Pauli operators.}
\label{fig:tetrontracking}
\end{figure}

\begin{figure*}
\centering
\def\svgwidth{0.97\linewidth}
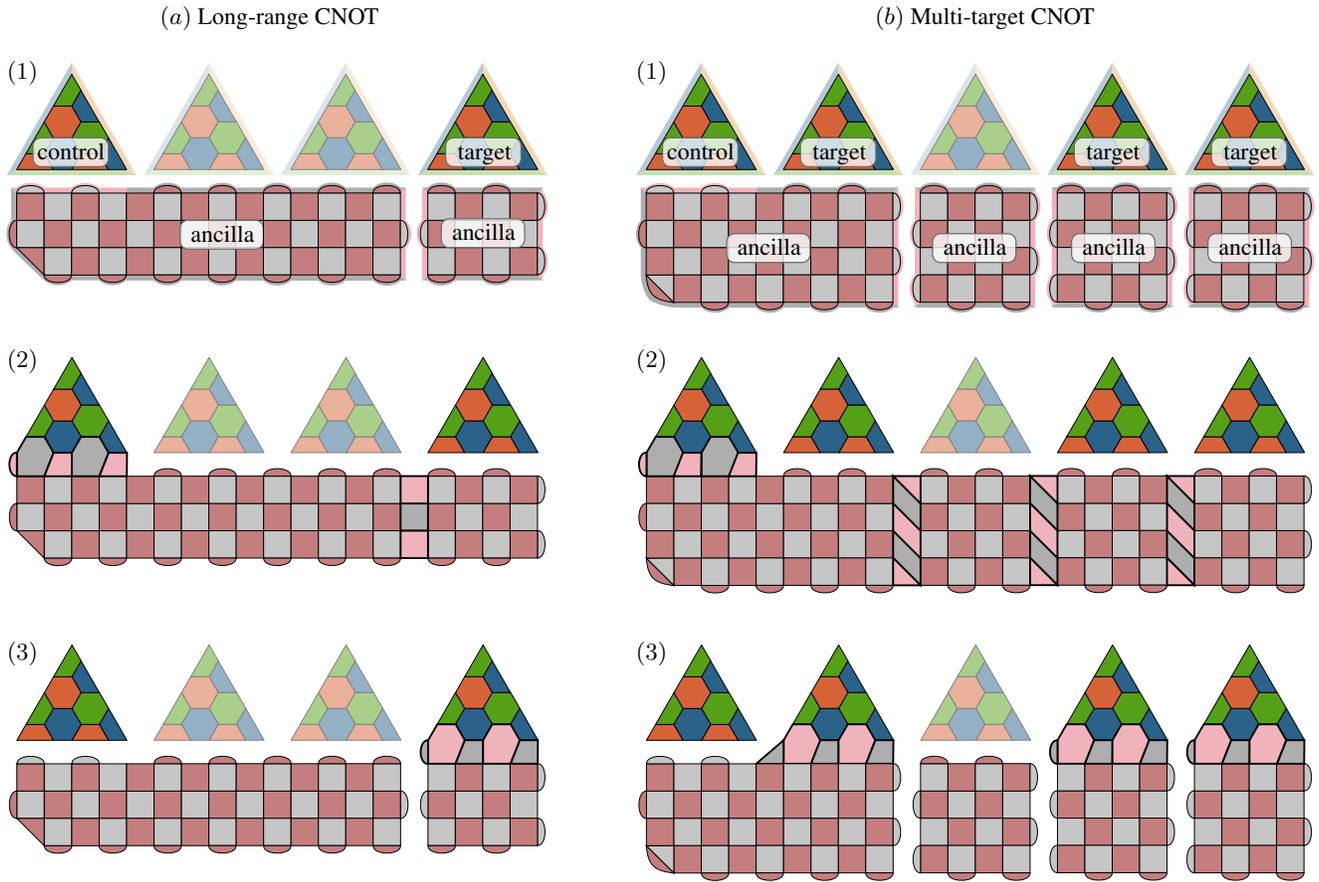
\caption{$(a)$ By simultaneously measuring the $ZZ$-parities between the control and two ancillas $(a2)$, one can use a long ancilla qubit for long-range CNOTs. The unused long ancilla is read out in the $\sigma_x$-basis. $(b)$ The same protocol can be used for long-range multi-target CNOTs, where multiple $ZZ$- and $XX$-parity measurements $(b2,b3)$ are carried out simultaneously. All protocols have a space overhead that scales with $\mathcal{O}(s \ln s)$ of the control-target separation $s$, and a time overhead that scales with $\mathcal{O}(\ln s)$.
}
\label{fig:cnot2}
\end{figure*}

We first discuss logical CNOT gates between neighboring color code qubits, see Fig.~\ref{fig:cnot1}b. The shape of the surface code qubit is chosen such that one $Z$-boundary is adjacent to the control qubit, and an $X$-boundary is next to the target qubit. In the first step $(b2)$, the $X$-stabilizers along the boundary with the control qubit are merged to form 6-qubit stabilizers (dark gray) and new $Z$-stabilizers (light purple) are introduced. While this is not evident from the figure, the boundary $Z$-stabilizers of the color code qubit remain unchanged. In the new configuration $(b2)$, all stabilizers commute, and the number of stabilizers has increased
by one, i.e., one bit of information is measured. As the gray boundary stabilizers are merely the product of the previously known boundary stabilizers, the only nontrivial measurement outcome is given by the purple boundary stabilizers. Since they contain each boundary qubit exactly once, their product is precisely the $ZZ$-parity between control and ancilla. Thus, lattice surgery provides a fault-tolerant logical parity measurement. Similarly, in the next step $(b3)$, lattice surgery merges the $Z$-stabilizers along the $X$-boundary of the surface code ancilla and a boundary of the target qubit. The product of gray boundary $X$-stabilizers yields the $XX$-parity. Finally, the surface code ancilla can be measured in the $\sigma_z$-basis by measuring all physical qubits and applying classical error correction, thereby completing the protocol of Fig.~\ref{fig:cnot1}a.

The role of Majorana tracking in this protocol is to appropriately update the composition of the stabilizers measured during lattice surgery. The protocol introduces $X$- and $Z$-stabilizers with non-coinciding support along the boundaries of the qubits. Thus, these stabilizers need to be appropriately updated by the tracking procedure described in Sec.~\ref{sec:logcliff}. For instance, a preceding $H$ gate on the control qubit in Fig.~\ref{fig:cnot1} would change the light purple boundary stabilizers in $(b1)$ from $\sigma_z^{\otimes 4}$ to $\sigma_x^{\otimes 2} \otimes \sigma_z^{\otimes 2}$. Accordingly, tracking would also change the dark gray boundary stabilizers to $\sigma_z^{\otimes 4} \otimes \sigma_x^{\otimes 2}$, and the red four-qubit boundary stabilizers of the control qubit to $\sigma_x^{\otimes 4}$. In Fig.~\ref{fig:tetrontracking}, we show the tetron tunnel coupling configurations used to measure these updated stabilizers for this particular example, but the procedure straightforwardly generalizes to all other possible cases. An update for the non-boundary color code stabilizers unaffected by lattice surgery can still be avoided, as their $X$- and $Z$-stabilizer support still coincides during the lattice surgery protocol.

Surface code ancillas are also useful for CNOT gates between color code qubits that are far away from each other. Lattice surgery can be used to measure the $ZZ$-parities between the control qubit and multiple ancilla qubits simultaneously~\cite{Horsman2012}, thereby initializing multiple ancillas at the same time. Consider the situation in Fig.~\ref{fig:cnot2}a, where the distance between two separated color code qubits is bridged by two surface code ancillas. Lattice surgery $(a2)$ can simultaneously measure the $ZZ$-parities between control and first ancilla and between both ancillas. This is equivalent to parity measurements between control and both ancillas, as the the $ZZ$-parity between control and second ancilla is given by the product of both measurements. Since the two $Z$-boundaries of the long ancilla are at opposite ends of the qubit, this lattice surgery step prepares an ancilla qubit adjacent to the distant target qubit for the next $XX$-parity measurement. Thus, this protocol yields a long-range CNOT gate between arbitrarily distant qubits with essentially the same time overhead as the nearest-neighbor CNOT. The unused long ancilla qubit cannot be discarded right away, as it is still entangled with the control qubit, but needs to be read out in the $\sigma_x$-basis with outcome $m$ by measuring all physical qubits in the $\sigma_x$-basis, leading to a $\sigma_z^m$ correction on the control qubit.

\subsection{Multi-target CNOTs for magic state distillation}

Clifford gates and physical $T$ gates are sufficient for universal quantum computing. One type of protocol using these ingredients for \textit{logical} $T$ gates is magic state distillation, whose precision scales with the protocol length. In such protocols, a physical magic state is initialized by applying a physical $T$ gate to a physical qubit in the $\ket{+}$-state. With tetrons~\cite{Karzig2016}, physical $T$ gates can be implemented via a measurement-based analog of the parity echo protocol introduced in Ref.~\cite{Karzig2015}. The resulting physical magic state is converted into a (faulty) logical magic state by code injection~\cite{Landahl2014,Litinski2017}, which requires only stabilizer measurements. Magic state distillation protocols convert many faulty magic states into fewer magic states with higher fidelity. Typically, these protocols rely on multi-target CNOT gates, i.e., CNOTs with one control qubit but multiple target qubits. For instance, the circuit corresponding to the 15-to-1 distillation protocol~\cite{Bravyi2005,Landahl2011} consists of 34 CNOT gates. But since many of these CNOTs have the same control qubit, the protocol actually requires only five \textit{multi-target} CNOTs.

Fortunately, lattice surgery can be used to implement multi-target CNOTs with the same time overhead as single CNOTs, as we show in Fig.~\ref{fig:cnot2}b. Here, lattice surgery measures the $ZZ$-parities between the control and each ancilla qubit $(b2)$. Therefore, each of the ancillas is treated like an ancilla qubit after step $(2)$ of the protocol in Fig.~\ref{fig:cnot1}a, but for multiple simultaneous CNOT protocols. After the $XX$-parity measurements between ancillas and their targets $(b3)$, the ancillas that were used for CNOTs are read out in the $\sigma_z$-basis, whereas the ancillas that were used to bridge long distances are read out in the $\sigma_x$-basis.

\begin{figure*}
\centering
\def\svgwidth{0.97\linewidth}
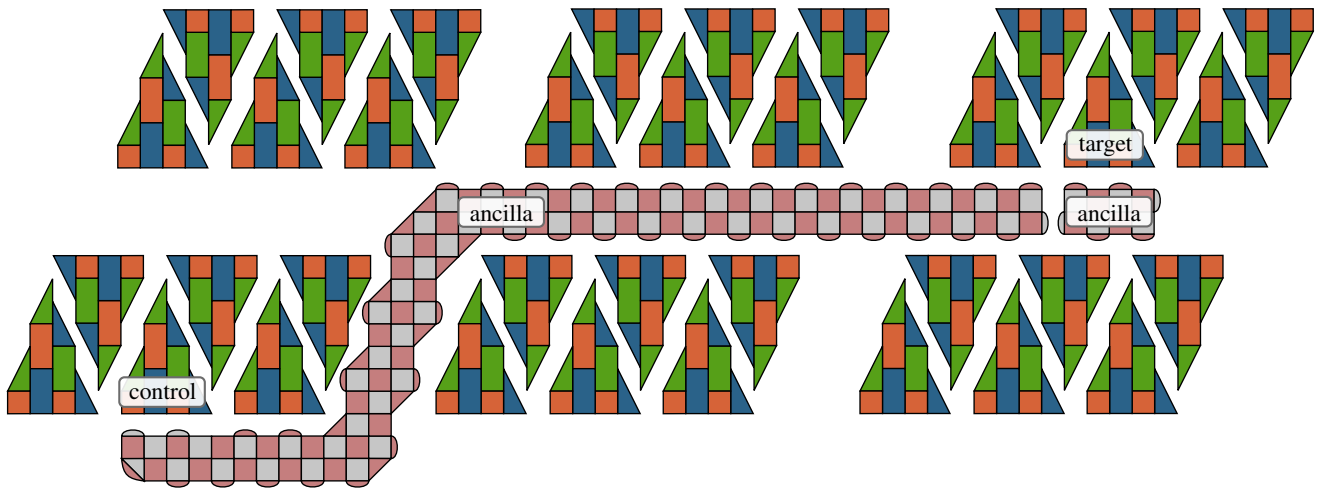
\caption{Example of a two-dimensional arrangement of color code qubits on a square lattice, and an example of surface code ancillas used for a long-range CNOT between distant qubits. For the next CNOT gate, these ancillas are discarded and the space between qubit blocks can be used to initialize different ancillas. The separation between blocks of color code qubits dictates the maximum code distance of the surface code ancillas, and influences the number of CNOT gates that can be performed in parallel. With larger separation, multiple ``lanes'' of ancilla qubits can fit between blocks, allowing for multiple overlapping multi-target CNOT gates.}
\label{fig:2dcnot}
\end{figure*}

In this way, we establish color-to-surface code lattice surgery as a useful tool for fault-tolerant long-range multi-target CNOT gates between color code qubits. Importantly, for a fixed code distance, the overhead of our protocol scales very favorably with the control-target separation $s$. As to the space overhead, strings of errors that connect the gray edges (or $X$-boundaries) of the long surface code qubit during the measurement of the $ZZ$-parities can lead to errors in the CNOT protocol. While these error strings are suppressed exponentially in the width of the surface code qubits, the number of possible strings grows linearly with their length. Thus, the width needs to increase with $\mathcal{O}(\ln s)$ in order to maintain the same CNOT accuracy. With a linearly growing length of the surface code qubits, the space overhead of the lattice surgery CNOT protocol is $\mathcal{O}(s \ln s)$.
For the time overhead, one needs to take the classical overhead of decoding and the effect of measurement errors during syndrome readout into account. There exist surface code decoders with a runtime of $\mathcal{O}(\ln s)$~\cite{Duclos2010}. The correction of measurement errors requires recording multiple rounds of syndrome extraction for one code cycle, depending on the measurement fidelity~\cite{Dennis2002}, effectively extending the code into a third time dimension. These ``time errors'' are suppressed exponentially with the number of measurement rounds, but the number of possible error strings increases linearly with $s$. Thus, similar to the space overhead, measurement errors increase the time overhead by $\mathcal{O}(\ln s)$, and the total time overhead is still only $\mathcal{O}(\ln s)$.

Note that the code distances (given by the width) of the ancilla qubits need not be as high as the code distance of the color code qubits, since the ancillas only need to survive for the few code cycles of the CNOT protocol, as opposed to data qubits that may need to survive for the entire quantum computation. In our example, the ancilla qubits have distances $d=3$, $d=4$, and $d=5$ in the protocols in Figs.~\ref{fig:cnot1} and \ref{fig:cnot2}. However, we expect that for most practical quantum computations, the entire space allocated for CNOT ancillas will be in use for different CNOTs essentially for the entire duration of the computation. Thus, for most practical purposes, the width of the ancilla qubits and the code distance of the color code qubits can be chosen to be equal (as in  Fig.~\ref{fig:cnot2}b), and the logarithmic scaling of the ancilla width can be ignored. With this approach, all parts of the code are protected against error strings of length $(d-1)/2$ during each code cycle. The logarithmic scaling merely implies that for a quantum computation involving $n$ logical qubits, the necessary code distance to reach a target error probability at the end of the quantum computation scales with $\mathcal{O}(\ln n)$.
Again, we point out that by identifying two copies of surface code qubits as one color code qubit~\cite{Kubica2015}, this CNOT protocol can be done entirely using color codes, as we show in Appendix \ref{app:surgery}. However, this uses more physical qubits than the surface code approach, and requires the measurement of 8-qubit stabilizers.

Both surface and color codes can be implemented on the square lattice of tetrons shown in Fig.~\ref{fig:tetrons}, since lattice surgery only requires the measurement of additional stabilizers, i.e., the measurement of 4-, 8-, and 12-Majorana operators. While our examples have illustrated logical qubits arranged on a line, this protocol can be straightforwardly extended to two-dimensional arrangements of logical qubits. One possible 2D arrangement of color code qubits is shown in Fig.~\ref{fig:2dcnot}, where qubits are arranged in blocks of six. The figure also shows two surface code ancilla qubits that can be used for a CNOT between distant color code qubits. In this way, lattice surgery can provide long-range communication between any two logical qubits with essentially constant time overhead.

\begin{figure*}
\centering
\def\svgwidth{\linewidth}
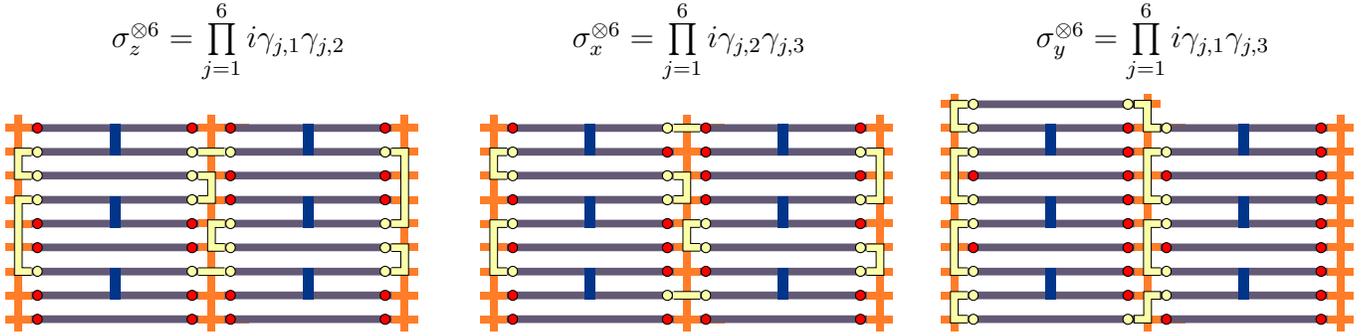
\caption{Tunnel coupling configurations for color code stabilizer measurements with tetrons that feature only one vertical semiconductor wire between tetrons, as in the architecture of Ref.~\cite{Karzig2016}.}
\label{fig:singlewire}
\end{figure*}

\section{Conclusion}
\label{sec:conclusion}

Current ideas for realizing a Majorana-based quantum computer rely on nanowire arrays such as networks of tetrons that only allow for local Majorana parity operator measurements and physical $T$ gates. Here, we have shown how Majorana-based qubits can be combined with color codes for universal fault-tolerant quantum computation without the need for moving Majoranas. In our architecture, logical single-qubit Clifford gates are implemented by Majorana tracking, which minimizes their overhead. Furthermore, we combine surface codes with color codes using surface-to-color code lattice surgery, which yields long-range multi-target CNOT gates with a time overhead that scales only with $\mathcal{O}(\ln s)$ of the distance $s$ between the control and target qubits, and a space overhead that scales with $\mathcal{O}(s \ln s)$. Moreover, this approach features a lower space overhead and lower-weight stabilizers compared to a purely color-code-based scheme.

Logical $T$ gates are the most expensive operation in this scheme, as they require magic state distillation. Their overhead can be reduced by improving the fidelity of physical $T$ gates, and by exploring faster distillation protocols and alternatives to magic state distillation. As to the concrete physical implementation, there are several proposals for architectures that implement the two operations required of nanowire arrays. Still, none of these architectures are particularly optimized towards error correction with color codes. Optimizing for fast stabilizer measurement, high measurement fidelity, and low physical error rate is crucial to ensure scalability.  Exploring efficient decoding schemes for color and surface code qubits can further reduce the classical overhead. 

What is more, our scheme can also be applied to non-topological architectures, such as superconducting qubits, albeit without the advantages of robust physical single-qubit Clifford gates and ancilla-free syndrome readout. For this reason, these architectures usually favor surface codes over color codes due to the easier four-qubit stabilizer measurements of surface codes compared to the weight-six stabilizers of color codes. Even though surface codes do not feature transversal single-qubit Clifford gates, they can still be used to implement single-qubit Clifford gates with zero time overhead~\cite{Hastings2015,Yoder2017,Litinski2017b} by encoding logical qubits in surface code twist defects~\cite{Bombin2010}.
Still, surface code qubits in this scheme suffer from a lack of easy $\sigma_y$ measurements, and require $\sim  d^2$ physical qubits for each logical qubit, while the color code approach discussed in this work only requires $\sim \frac{3}{4}d^2$ physical qubits to achieve the same code distance $d$.

We note that the twist-based triangle codes presented in Ref.~\cite{Yoder2017} also feature a space overhead of $\sim \frac{3}{4}d^2$ and manage to implement easy $\sigma_y$ measurements, but they encode the logical $\sigma_x$, $\sigma_y$, and $\sigma_z$ information in each of the three sides of the triangles separately. Therefore, these qubits cannot be packed as densely as the color code qubits in Fig.~\ref{fig:2dcnot}, since all three sides of each triangle need to be accessible by lattice surgery. This either implies an increased space overhead by requiring some free space as padding around the triangles, or it introduces a time overhead for logical single-qubit Clifford gates by requiring code operations for the reorientation of triangle qubits.

The spatial overhead of the color code scheme, on the other hand, can be reduced even further to $\sim \frac{1}{2}d^2$ by using 4.8.8 color codes~\cite{Landahl2011} instead of the 6.6.6 color codes discussed in this work. However, this comes at the price of higher-weight stabilizer, as 4.8.8 color codes feature eight-qubit stabilizers, instead of just six-qubit stabilizers. If higher-weight stabilizers are not significantly more difficult to measure, which could hold true for Majorana-based qubits, it is advantageous to use the color-code-based scheme instead of a pure surface code architecture in order to reduce the overhead of fault-tolerant quantum computing.


\section*{Acknowledgments}

We thank C.~E.~Bardyn, J.\ Eisert, T.\ Karzig, M.~S.~Kesselring, and S.\ Plugge for illuminating discussions. This work has been supported by the Deutsche Forschungsgemeinschaft (Bonn) within the network CRC TR 183.

\appendix

\section{Stabilizer measurements with single-wire tetrons}
\label{app:singlewire}

Here, we show how the measurement of the color code stabilizers shown in Fig.~\ref{fig:tetrons} can be implemented in a network of tetrons that features only one vertical semiconductor wire between tetrons, which is the architecture discussed in Ref.~\cite{Karzig2016}. The corresponding tunnel coupling configurations in this architectures are shown in Fig.~\ref{fig:singlewire}.

\begin{figure}[b]
\centering
\def\svgwidth{\linewidth}
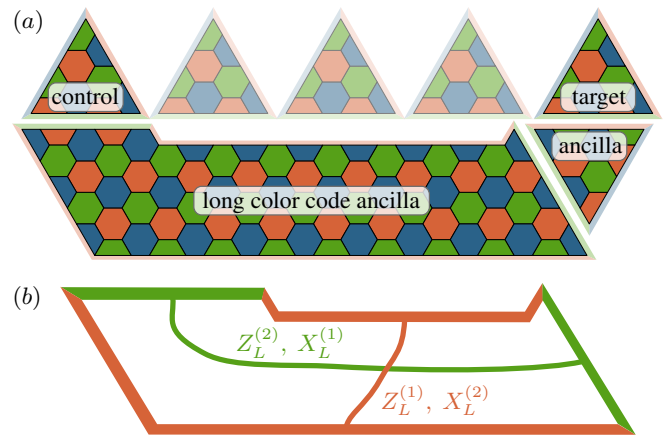
\caption{$(a)$ Qubit arrangement for long-range CNOTs by lattice surgery using a color code ancilla. $(b)$ The long color code qubit used in this protocol actually encodes two logical qubits, as it has two red boundaries and two green boundaries. Red-to-red strings define the logical operators $Z_L^{(1)}$ and $X_L^{(2)}$, while green-to-green strings are the operators $Z_L^{(2)}$ and $X_L^{(1)}$.}
\label{fig:colorcodecnot}
\end{figure}

In the configuration of Fig.~\ref{fig:tetrons} with two vertical wires, all stabilizers of the same color can be measured simultaneously. This is no longer the case in Fig.~\ref{fig:singlewire}, as these measurements use a vertical wire or a coherent link of a neighboring six-qubit block. In particular, the $\sigma_z^{\otimes 6}$- and $\sigma_x^{\otimes 6}$-measurements overlap with their left (or right) neighbors, whereas the $\sigma_y^{\otimes 6}$-measurement overlaps with the upper neighbor. Therefore, syndrome extraction requires \textit{two} measurement rounds for the measurement of each stabilizer type, as opposed to just one.

\section{Lattice surgery with color code ancillas}
\label{app:surgery}

The long-range CNOT protocol can also be done using color code ancillas. Here, the long surface code qubit is replaced by a long color code qubit, see Fig.~\ref{fig:colorcodecnot}a. This is a color code qubit with two red and two green boundaries, as shown in Fig.~\ref{fig:colorcodecnot}b. It is defined by $n$ physical qubits and $n-2$ stabilizers, and therefore encodes 2 logical qubits. However, since the code distance of such a qubit is always even, the support of the $X_L$ and $Z_L$ operators of the individual logical qubits cannot coincide. Instead, strings of Pauli operators that connect two green boundaries encode the operators $X_L^{(1)}$ and $Z_L^{(2)}$, and red-to-red strings are $X_L^{(2)}$ and $Z_L^{(1)}$, where the superscript labels the logical qubit.

In this way, long color code qubits are equivalent to two surface code qubits on top of each other with a relative rotation of 90 degrees. Therefore, the long-range CNOT protocol is the same as for surface code ancillas, but only uses one of the two encoded logical qubits. This redundancy is also manifested in a higher number of physical qubits than in surface code ancillas. Moreover, lattice surgery between color codes requires 8-qubit stabilizer measurements.

%

%
%

\bibliographystyle{apsrev4-1mod}
\bibliography{biblio}

\end{document}